\documentclass[aps,prb,onecolumn,groupedaddress,showpacs]{revtex4-1}

\usepackage{dcolumn}
\usepackage{graphicx,amssymb,amsmath}
\usepackage{bm}
\usepackage{natbib}
\usepackage{epstopdf}
\begin{document}

\title{Distribution Function for $n \ge g$ Quantum Particles  }
\author{Shimul Akhanjee}
\affiliation{ akhanjee.shimul@gmail.com }

\date{\today}

\begin{abstract}
A new quantum mechanical distribution function $n^I(\varepsilon)$, is derived for the condition $n \ge g$, where in contrast to the exclusion principle $n \le g$ for fermions, each energy state must be populated by at least one particle. Although the particles share many features with bosons, the anomalous behavior of $n^I(\varepsilon)$ precludes Bose-Einstein condensation (BEC) due to the required occupancy of the excited states, which creates a permanently pressurized background at $T=0$, similar to the degeneracy pressure of fermions. An exhaustive classification scheme is presented for both distinguishable and indistinguishable, particles and energy levels based on Richard Stanley's twelvefold way in combinatorics. 
\end{abstract}

\maketitle

\section{Introduction}
The statistical distribution function, $n(\varepsilon)$ for identical particles has  been an essential component of quantum mechanics. Historically, the behavior of $n(\varepsilon)$ has been over-determined by key experimental facts in a wide variety of physical systems such as the blackbody spectrum, semiconductor heterostructures, astrophysical spectroscopic measurements, low temperature $T$, and condensed matter systems\cite{kittel,astro,RevModPhys.73.307,leggettbook}. Theoretical approaches converge, from the grand-canonical ensemble to the micro-canonical ensemble, as well as the more mathematically rigorous Darwin-Fowler method of mean values\cite{bose1924,einstein1924,fermi1926,dirac1926,arnaud1999,darwin1922}. 

Consider a system within the microcanonical ensemble having a fixed number of particles $N = \sum_j {n_j}$, total energy $U = \sum_j {\varepsilon_jn_j}$, and volume $V$\cite{schwabl}. One can make use of the mathematical structures found in combinatorial counting problems, particularly the number of ways that one can distribute a specified number of balls into a fixed number of boxes as shown in Table 1, known as Richard Stanley's twelvefold way\cite{stanleybook}. For quantum systems in particular there are only three possible arrangements of the identical balls, of which represent indistinguishable particles, into the labeled boxes that play the role of distinguishable energy states. I will introduce the new case of the second row, third column of Table 1.

\begin{center}
\begin{tabular}{ |p{2cm}||p{2cm}|p{2cm}|p{2cm}|  }
\hline
\multicolumn{4}{|c|}{\bf{Table 1: The Twelvefold Way - How many ways can $\bf {n}$ balls be sorted into $\bf{g}$ boxes?} }\\
\multicolumn{4}{|c|}{${n\brace g}$ - Stirling numbers of the 2nd kind  }\\
\multicolumn{4}{|c|}{$p_{\le g}(n)$ - integer partitions of $n$ into at most $g$ parts }\\
\multicolumn{4}{|c|}{$p_{ g}(n)$ - integer partitions of $n$ into exactly $g$ parts }\\
\hline
Ball and Box Set & Arbitrary (Any Sorting) & Injective (Maximum 1 ball per box) & Surjective (Minimum 1 ball per box)\\
 \hline\hline
 Distinct Balls \\Distinct Boxes   & $g^n$    &$\frac{g!}{(g-n)!}$& $g! {n\brace g}$\\
 \hline
 Identical Balls \\ Distinct Boxes &   ${g+n-1}\choose{n}$  & ${g}\choose{n}$   &${n-1}\choose{g-1}$\\
 \hline
Distinct Balls\\Identical Boxes & ${\sum}_{j=0}^g {n\brace j}$ & 1 if $n \le g$&  $n \brace g $\\
 \hline
 Identical Balls\\Identical Boxes    &$p_{\le g}(n)$ & 1 if $n \le g$&  $p_{ g}(n)$\\

 \hline
\end{tabular}
\end{center}

\section{Unrestricted sorting of $n$ and $g$} Starting with second row, first column of Table 1, the microstate configuration of bosons can be constructed from the distinct orderings of $g_j -1$ lines and $n_j$ circles as shown in Fig.\ref{fig:figure1}(a)
\begin{equation} 
t^{B}_{j}= { {g_j+n_j -1}\choose{n_j} }= \frac{(g_j+n_j -1)!}{n_{j}!(g_j-1)!}
\label{BEmicro}
\end{equation}
where the standard manipulations lead to the Bose-Einstein distribution,
\begin{equation}
n^{B}_{j}(\varepsilon)= \frac{g_j }{ \mathrm{e}^ { (\varepsilon_j - \mu)/ (k_B T) }   -1 }
\label{bdist}
\end{equation}
 
\section{$n_j \le g_j$ - The exclusion principle}
\begin{figure}
\centerline{\includegraphics[width=2.5in]{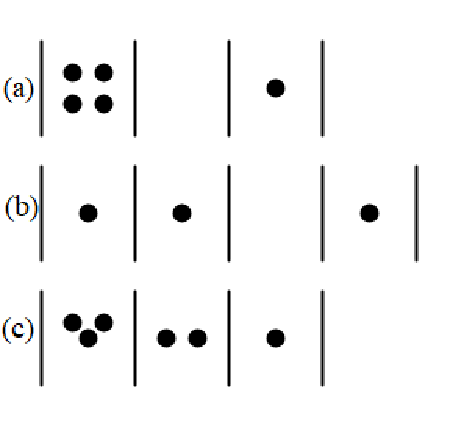}}
\caption{Typical configurations of the three combinatorially distinct possibilities for identical particles distributed into distinguishable states. (a) Unrestricted sorting, allowing for more than one particle in a state, in addition to empty states. (b) The exclusion principle, $n_j \le g_j$, with no more than one particle per state, and allowing empty states. (c) The new case introduced here: $n_j \ge g_j$, where all sub-states must be occupied by at least one particle, while no upper bound is imposed.} 
\label{fig:figure1}
\end{figure}

. Next, by examining the second row, second column of Table 1 the resulting distribution represents fermions. This particular occupancy of the energy levels is depicted in Fig.\ref{fig:figure1}(b), where this scenario implies that only one particle can occupy a sub-state of $g$, 
\begin{equation} 
t^{F}_{j}= { {g_j}\choose{n_j} }= \frac{g_j!}{n_{j}!(g_j-n_j)!  }
\end{equation}
This yields the Fermi-Dirac distribution,
\begin{equation}
n^{F}_{j}(\varepsilon)= \frac{g_j }{ \mathrm{e}^ { (\varepsilon_j - \mu)/ (k_B T) }   +1 }
\label{fdist}
\end{equation}

\section{$n_j \ge g_j$ - The inclusion constraint}. 

The two preceding scenarios are not exhaustive. The primary purpose of this paper is to demonstrate the combinatorially distinct possibility of the second row, third column of Table 1. The inclusion principle introduced here, is a new case of quantum statistics.  A single particle is attached to every positive energy level, requiring $n_j \ge g_j$, such that no positive energy level is vacant as shown in Fig.\ref{fig:figure1} (c). Although, a similar occupation of the excited states might be possible in the classical limit $k_B T \gg \varepsilon_j$, where the phase-space density (number of particles per quantum state) is very high, here it is not assumed that this condition is generated from external factors. Rather, the level occupancy is presupposed as an intrinsic property of the particles. Therefore, the microstate configuration of interest can be adapted from the surjective case for identical balls in distinct boxes\cite{stanleybook},
\begin{equation} 
t^{I}_{j}= { {n_j-1}\choose{g_j-1} }= \frac{(n_j-1)!}{(g_{j}-1)!(n_j-g_j)!}
\label{imicro}
\end{equation}

For large values of $n$ and $g$, the total number of microstates becomes a product,
$t_T \approx \prod_j{\frac{n_j !}{g_j !(n_j-g_j)!  }}$ and after the use of Stirling's approximation: $\ln N! \approx N \ln N -N$, the entropy $S =k_B\ln{t_T}$  can be expressed as.
\begin{equation}
S=k_B \sum_j{}n_j \ln{n_j } -g_j \ln{g_j }-(n_j -g_j)\ln{(n_j -g_j)} 
\end{equation}
Next, we develop the condition for an entropy maximum. Derivatives are taken with respect to $n_j$. The macrostate conditions $\mathrm{d}N = \sum_j {\mathrm{d} n_j} =0$ and $\mathrm{d} U =  \sum_j {\varepsilon_j \mathrm{d}n_j} =0$ are enforced with Lagrange multipliers $\alpha$ and $\beta$, 
\begin{equation}
\frac{\mathrm{d}S}{\mathrm{d}n_j} = \sum_j \ln \left( \frac{n_j}{n_j-g_j} \right)-\alpha -\beta\varepsilon_j =0
\end{equation}
Evidently, a dimensional analysis of the thermodynamic potential $\mathrm{d}U=\frac{1}{\beta}\mathrm{d}S-\frac{\alpha}{\beta}\mathrm{d}N$, reveals the correspondence with  $k_B T = 1/\beta$ and the chemical potential $\mu = -\alpha/\beta$. After solving for $n_j$, the final expression becomes,
\begin{equation}
\boxed{
n^{I}_j(\varepsilon_j) =  \frac{g_j \mathrm{e}^{\beta(\varepsilon_j-\mu)} }{\mathrm{e}^ {\beta(\varepsilon_j-\mu)} - 1}}
\label{distfinal}
\end{equation}

Consider a three dimensional, non-interacting gas of these particles with energy $\varepsilon_{p}=\frac{p^2}{2m}$. Apparently, the fixed background of excited states will have important thermodynamic consequences.  It may be tempting to assign a spin $s$ to such quantum particles, where in the absence of a magnetic field, the degeneracy factor is $g=2s+1$. However, no assumptions about the permutation symmetry under the exchange of two particles should be made without a more rigorous development of the Fock space. Naively, in the number occupancy basis the eigenstates are given by,

\begin{equation}
\left|n_{p_1},n_{p_2},\cdots n_{p_N}\right>\propto\left(\left|\psi_{p_1}\right>\right)^{n_{p_1}}\left(\left|\psi_{p_2}\right>\right)^{n_{p_2}},\cdots\left(\left|\psi_{p_N}\right>\right)^{n_{p_N}}
\end{equation}
In order to enforce the inclusion principle mathematically, the state $\left|\Gamma\right\rangle $ defined below,
\begin{equation}
\left|\Gamma\right\rangle=\left|1,1,1,\cdots\right>
\end{equation}
must vanish after applying annihilation operator $a_p \left|\Gamma\right\rangle=0 $, for all values of $p$. This condition will definitely have consequences for the structure of the wavefunctions. On the other hand, for the usual bosonic case, a ``simple" BEC transition occurs where the $\vec{p}=0$ state is macroscopically occupied at $T=0$\cite{einstein1924,leggettbook},
\begin{equation}
\left|BEC\right>=\left|N_{_{}},0,0,\cdots\right>\propto \left(a_{p=0}^{\dag}\right)^{N-1}\prod_{p\ne0}a_p\left|\Gamma\right>=0
\end{equation}
Thus, the $\left|BEC\right>$ state is forbidden because of the condition $a_p \left|\Gamma\right\rangle=0 $.

Alternatively, one can derive Eq.(\ref{distfinal}) exactly, with no approximations by applying the grand canonical ensemble\citep{kardar}. The conventional approach takes on summations over each occupation number $n_p$, with allowed values: $[0,1]$ for fermions and $[0,1,2,\dots]$ for bosons. For the new case considered here: $[1,2,\dots]$ or $n_p \ne 0$. The grand partition function for that system becomes,
\begin{equation}
\mathcal{Z}_G=\underset{p}{\prod}\underset{n_{p}\ne0}{\sum}e^{-\beta(\varepsilon_p-\mu)n_p}=\underset{p}{\prod}\frac{e^{-\beta(\varepsilon_p-\mu)}}{1-e^{-\beta(\varepsilon_p-\mu)}}
\label{grandpart}
\end{equation}
where the geometric series converges only if $e^{-\beta\left(\varepsilon-\mu\right)} <1 $, which is true for $\mu <0$. Therefore, the average particle number for a single particle sub-state: $\mathcal{Z}_G=\underset{p}{\prod}\mathcal{Z}_p$ becomes, 
\begin{equation}
\left\langle n_{p}\right\rangle=k_BT\frac{1}{\mathcal{Z}_p}\left(\frac{\partial\mathcal{Z}_p}{\partial\mu}\right)_{V,T}= \frac{ \mathrm{e}^{\beta(\varepsilon_p-\mu)} }{\mathrm{e}^ {\beta(\varepsilon_p-\mu)} - 1}
\label{grandpot}
\end{equation}
Hence, Eq.(\ref{distfinal}) that was derived earlier within the microcanonical ensemble is also confirmed by the grand canonical ensemble approach, given that for a single particle level, $g\left\langle n_{p}\right\rangle= n^{I}(\varepsilon_p)$. As expected, the particle number variance $\sigma_N^2=k_BT\left(\frac{\partial\left\langle n_p\right\rangle}{\partial \mu}\right)_{V,T}\sim\frac{(k_BT)^2}{(\varepsilon-\mu)^2}$, retains the bosonic form at $k_B T \gg \varepsilon_j$ since the fixed occupancy of excited states should not significantly contribute to $\sigma_N$.
\begin{figure}
\centerline{\includegraphics[width=3.7in]{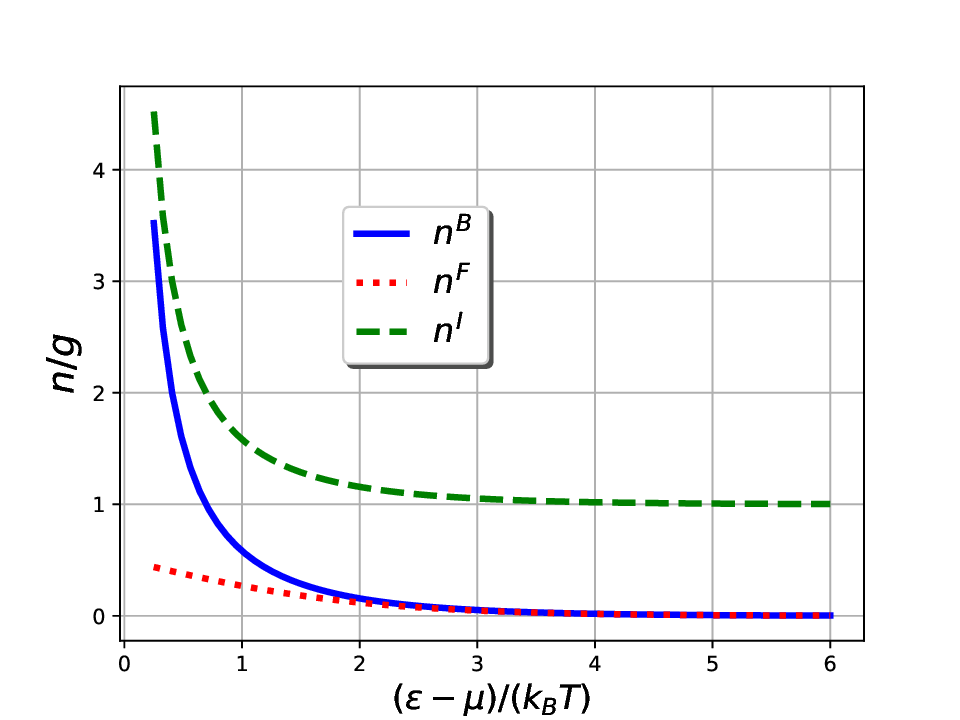}}
\caption{Comparison of the 3 quantum distributions. Unlike $(n^B)/g$ and $(n^F)/g$, which decay to zero at $(\varepsilon-\mu)\gg k_B T$, $(n^B)/g$ saturates at unity.} 
\label{fig:figure2}
\end{figure}

\section{Thermodynamics}. It is useful to express Eq.(\ref{distfinal}) as,
\begin{equation}
n_j^I(\varepsilon_j) = n_{j}^B(\varepsilon_j)+g
\end{equation}
An analysis of various thermodynamic quantities in terms of different components will help to elucidate the physical properties of the system. The $\vec{p}=0$ contribution should be separated and treated carefully, especially as $T \rightarrow 0$, or where the fugacity $z=e^{\mu \beta} \rightarrow 1$. In the limit of large $V$, and constant specific volume $v=V/N$, the sums over discrete energy levels for the excited states can be replaced by integrals over $g{\sum}_{\vec{p} \ne 0}{\cdots}=g\frac{V}{\left(2 \pi\hbar\right)^{3}}\int_{}^{}d^3p$. The average particle number becomes the sum of three terms, 
 
\begin{equation}
N= \sum_{p}{n^I(\varepsilon_p)} = N_0(z) + N_1(z) + N_2
\end{equation}
The first term describes the number of particles occupying the ground state energy,
\begin{equation}
N_0(z)=n^I(\varepsilon_{p=0})=\frac{g}{1-z}
\end{equation}
of which, diverges at $z=1$. Moreover, $N_1(z)$ and $N_2$ account for the excited states of the $n^B$ term and occupied background respectively,
\begin{equation}
N_1(z)=\frac{V}{\left(2\pi\hbar\right)^{3}}\int{n^B\left(\varepsilon_p\right)d^3p}=\frac{gNv}{\lambda^3}f_{3/2}^{+}(z)
\end{equation}
\begin{equation}
N_2=\frac{gVm^{3/2}}{\sqrt{2}\pi^{2}\hbar^{3}}\int_0^{\Omega_{}}\varepsilon^{1/2}d\varepsilon=\frac{4}{3\sqrt{\pi}}\frac{gNv}{\lambda^3}\left(\frac{\Omega}{k_B T}\right)^{3/2}
\label{equN2}
\end{equation}
where the thermal wavelength is defined by $\lambda = \hbar \sqrt{(2\pi)/(mk_BT)}$ and the generalized $\zeta$ function is defined by $f_{\nu}^{+}(z)=\frac{1}{\left(\nu-1\right)!}\int_0^{\infty}\frac{x^{\nu-1}}{z^{-1}e^{x}-1}dx{\frac{}{}}$\cite{zwillinger}.
Upon inspection, it is clear that Eq.(\ref{equN2}) would seem problematic since it diverges at the upper limit of integration. Consequently, a high energy cutoff $\Omega$ should ensure finite results.

In order to determine whether the system transitions into a ``simple" BEC state, it is necessary to study the behavior of the condensate fraction,
\begin{equation}
\nu_0=\underset{{N}\rightarrow\infty}{\mathrm{\lim}}N_0(z)/(N(z))
\end{equation}
Since $N_1(z=1)$ has a limiting value in the conventional BEC transition, $N_1(z=1)\propto T^{3/2}$, and therefore the number of excited states arising from $N_1(z=1)$ vanishes at low temperatures. However, since $N_2$ is independent of both $T$ and $z$, there cannot be a macroscopic occupation of the ground state unless $\Omega$ is sufficiently small. Thus, $\nu_0$ can never reach the value of $1$ and a complete BEC transition is not possible. This should be apparent from the outset since a significant fraction of the excited states are permanently occupied and can never move into the ground state energy.

The pressure $P$ can be determined from the grand potential $\Phi$, starting with Eq.(\ref{grandpart}):
\begin{equation}
\begin{aligned}
P&=-\frac{\Phi}{V}=\frac{1}{V\beta}\mathrm{\ln}\left(\mathcal{Z}_G\right)=\frac{1}{V\beta}{\sum_p\mathrm{\ln}\left(\frac{e^{-\beta(\varepsilon_p-\mu)}}{1-e^{-\beta(\varepsilon_p-\mu)}}\right)}\\
&=\frac{gk_BT}{\lambda^3}f^{+}_{5/2}(z)+\frac{4}{5\sqrt{\pi}}\frac{g}{\lambda^3}\frac{\Omega^{5/2}}{(k_B T)^{3/2}}-\frac{\mu N_2}{V}
\end{aligned}
\label{pressure}
\end{equation}

The common wisdom suggests that as the distribution function of a quantum gas flattens, which generally occurs as $T$ increases, then $P$ increases, reflecting a shift away from quantum effects and toward classical ideal gas behavior.
Furthermore, higher excited state occupation results in a higher average kinetic energy of the gas, which translates directly to higher $P$, since it is associated with particle collisions and their average momentum transfer. When approaching the  $T\to0$ limit of Eq.(\ref{pressure}), $P\sim \Omega^{5/2}$, which is similar to the degeneracy pressure of fermions $P_F=\frac{2}{5}\varepsilon_F\frac{N}{V}$, where the cutoff $\Omega$ is analogous to the Fermi level, $\varepsilon_F$. 

\section{Conclusion}
After examining the second row of Table. 1, an exhaustive analysis of the possible distribution functions for identical particles populating distinguishable energy levels has been undertaken. However, the fourth row suggests the possibility of indistinguishable energy levels. The Gibbs paradox points out that from from a classical standpoint, the non-extensivity of the entropy arises due to the neglect of the factor $1/N!$ when over-counting configurations of the partition function for identical particles\cite{schwabl}. It would be an interesting endeavor to study the consequences of the microstates being constructed from different integer partitions of $n$ into $g$ parts. New paradoxical inconsistencies could arise from enforcing the mathematical conditions that account for identical energy states.

To conclude, a classification scheme for identical particles has been developed by applying important results from enumerative combinatorics, namely the twelvefold way of $n$ balls sorted into $g$ boxes. The distribution function, $n^I(\varepsilon)$ for $n \ge g$ quantum particles has, for the first time, been derived exactly from within the microcanonical and the grand canonical ensembles. At first glance, many of its features are similar to bosons however a ``simple", non-fragmented BEC state is prevented and the system exhibits a $T=0$ pressure that is energy cutoff dependent. In other words, the system shares features of both fermions and bosons. Such particles could have tremendous implications in various high energy, astrophysical and cosmological theories, specifically dark matter candidates. Unlike ordinary matter, dark matter is ``collisionless" under normal conditions, meaning dark matter particles rarely interact with each other or with regular matter in a way that would create traditional pressure\cite{darkreview}. Furthermore, the inclusion principle could possibly explain some anomalies in astrophysical observations. For example, in galactic halos, a non-fermionic $T=0$ pressure present in dark matter could act as a stabilizing factor against gravitational collapse. In galaxy clusters, this helps the dark matter halo retain its shape, size, and density profile\cite{darkprimer}.

\section{Conflicts of Interest}

The author has no conflicts of interest to declare that are relevant to the contents of this article.

\end{document}